\documentclass[12pt]{article}

\usepackage{scicite}

\usepackage{times}
\usepackage[left]{lineno}
\usepackage{graphicx}
\usepackage{longtable}
\usepackage{setspace}
\usepackage{multirow}
\usepackage{xspace}
\usepackage{color}
\usepackage{amssymb}
\usepackage{ulem}

\newenvironment{sciabstract}{%
\begin{quote} \bf}
{\end{quote}}

\newcounter{lastnote}

\newcommand{\taumix}{{\ensuremath{\tau_\mathrm{mix}}}\xspace}

\title{Measurement of $^{19}$F($p$,\,$\gamma$)$^{20}$Ne reaction suggests CNO break-out in first stars}

\author
{Liyong Zhang,$^{1}$ Jianjun He,$^{1,*}$ Richard J. deBoer,$^{2}$ Michael Wiescher,$^{2,*}$ \\
Alexander Heger,$^{3}$ Daid Kahl,$^{4}$ Jun Su,$^{1}$ Daniel Odell,$^{5}$ Yinji Chen,$^{1}$ \\Xinyue Li,$^{1}$ Jianguo Wang,$^{6}$ Long Zhang,$^{7}$ Fuqiang Cao,$^{7}$ Hao Zhang,$^{1}$ \\
Zhicheng Zhang,$^{8}$ Xinzhi Jiang,$^{1}$ Luohuan Wang,$^{1}$ 
Ziming Li,$^{1}$ Luyang Song,$^{1}$ \\Hongwei Zhao,$^{6}$ Liangting Sun,$^{6}$ Qi Wu,$^{6}$  
Jiaqing Li,$^{6}$ Baoqun Cui,$^{7}$ Lihua Chen,$^{7}$ \\Ruigang  Ma,$^{7}$ Ertao Li,$^{8}$ 
Gang Lian,$^{7}$ Yaode Sheng,$^{1}$ Zhihong Li,$^{7}$ Bing Guo,$^{7}$ \\Xiaohong Zhou,$^{6}$ Yuhu Zhang,$^{6}$ 
Hushan Xu,$^{6}$ Jianping Cheng,$^{1}$ Weiping Liu,$^{7,*}$ \\
\normalsize{$^{1}$Key Laboratory of Beam Technology of Ministry of Education,}\\
\normalsize{College of Nuclear Science and Technology, Beijing Normal University, Beijing 100875, China}\\
\normalsize{$^{2}$Nuclear Science Laboratory, University of Notre Dame, Notre Dame, Indiana 46556, USA}\\
\normalsize{$^{3}$School of Physics and Astronomy, Monash University, Victoria 3800, Australia}\\
\normalsize{$^{4}$Extreme Light Infrastructure -- Nuclear Physics, Horia Hulubei National Institute for R\&D}\\
\normalsize{in Physics and Nuclear Engineering (IFIN-HH), Bucharest-M\u{a}gurele 077125, Romania}\\
\normalsize{$^{5}$Institute of Nuclear and Particle Physics and Department of Physics and Astronomy,}\\ \normalsize{Ohio University, Athens, Ohio 45701, USA}\\
\normalsize{$^{6}$Institute of Modern Physics, Chinese Academy of Sciences, Lanzhou 730000, China}\\
\normalsize{$^{7}$China Institute of Atomic Energy, P. O. Box 275(1), Beijing 102413, China}\\
\normalsize{$^{8}$College of Physics and Optoelectronic Engineering, Shenzhen University, Shenzhen 518060, China}\\
\\
\normalsize{$^\ast$Corresponding author. E-mail: hejianjun@bnu.edu.cn, michael.c.wiescher.1@nd.edu, wpliu@ciae.ac.cn}
}

\date{}

\begin{document}


\baselineskip24pt
%\baselineskip12pt


\maketitle


\begin{sciabstract}
The origin of calcium production in the first stars (Pop III stars), which formed out of the primordial matter of the Big Bang, and their fates, remain most fascinating mysteries in astrophysics.  Advanced nuclear burning and supernovae were thought to be the dominant source of the Ca production seen in all stars.  Here we report on a qualitatively different path to Ca production through break-out from the ``warm” carbon-nitrogen-oxygen (CNO) cycle.  We extend direct measurement of the $^{19}$F($p$,\,$\gamma$)$^{20}$Ne break-out reaction down to an unprecedentedly low energy point of 186~keV and discover a key resonance at 225~keV.  In the domain of astrophysical interest, at around 0.1~giga kelvin, this thermonuclear $^{19}$F($p$,\,$\gamma$)$^{20}$Ne rate is up to a factor of 7.4 larger than the previous recommended rate.  Our stellar models show a stronger break-out during stellar hydrogen burning than thought before, and may reveal the nature of Ca production in Pop III stars imprinted on the oldest known ultra-iron poor star, SMSS0313-6708. 
This result from the China Jinping Underground Laboratory, the deepest laboratory in the world, offering an environment with extremely low cosmic-ray induced background, has far-reaching implications on our understanding of how the first stars evolve and die.  Our rate showcases the impact that faint Pop III star supernovae can have on the nucleosynthesis observed in the oldest known stars and first galaxies, key mission targets of the James Webb Space Telescope. 

\end{sciabstract}

\noindent 
Stars are the nuclear forges of the cosmos, responsible for the creation of most elements heavier than helium in the Universe. 
Some of these elements are created in the hearts of stars over the course of billions of years, whereas others are formed in just a few seconds during the explosive deaths of massive stars. These heavy elements play an important role in the universe, allowing for the formation of complex molecules and dust which facilitate the cooling and condensation of molecular clouds, aiding the formation of new stars like our Sun. The first generation of stars, called Population III, Pop III stars, or primordial stars, formed from the pristine matter left by the Big Bang, thus play a special role in seeding the universe with the first heavy elements and creating suitable conditions for future generations of stars and galaxies. 

Every star, regardless of its mass, spends the majority of its life quiescently fusing hydrogen into helium in its core through two primary mechanisms: the $p$-$p$ chains and the catalytic carbon-nitrogen-oxygen (CNO) cycles~\cite{b2fh,rol88,ade11}. Which mechanism dominates hydrogen burning is determined by the temperature in the core of a star.  In stars with initial masses less than $\sim$1.2~solar masses ($\mathrm{M}_\odot$), with relatively cool cores ($T\leq0.02$~GK), the $p$-$p$ chains dominate the hydrogen fusion, whereas in stars with higher initial masses and hotter cores, the CNO cycles take over. As a catalytic reaction, the total number of CNO nuclei remains constant, unless a breakout reaction sequence causes a leakage toward the NeNa mass region, or if temperature and density are high enough to forge new carbon by the triple-alpha (3$\alpha$) process. The latter two occur in primordial massive stars. The only reaction that can potentially remove the catalytic material from the cycle at lower temperatures is the fusion of $^{19}$F with a proton to form $^{20}$Ne, denoted $^{19}$F($p$,~$\gamma$)$^{20}$Ne~\cite{wie99}. 
Previously, this reaction was thought to be rather weak compared to the competing $^{19}$F($p$,~$\alpha$)$^{16}$O reaction, so most of the $^{19}$F produced by the CNO cycle would be recycled back into $^{16}$O, with no significant chemical abundance changes~\cite{arn95}. 

The most metal-poor stars observed in our Milky Way's halo today display the diluted nucleosynthetic signatures resulting from Pop III stars that preceded them~\cite{fre15}.  Keller~\textit{et al.}~\cite{kel14} discovered one of the oldest known stars in the Universe, SMSS0313-6708, and, based on the stellar models by Heger \& Woosely~\cite{hw10}, suggested that a CNO breakout during hydrogen burning is the source of calcium (Ca) production, reporting [Ca/H] = --7.2~\cite{kel14}.  Takahashi \textit{et al.}~\cite{tak14} also cited such a breakout as the Ca production mechanism for the stars HE 1327-2326 and HE 0107-5240, with [Ca/H] = --5.3 and --5.13, respectively. Pop III stars begin their lives with primordial Big-Bang composition and contract until the central temperature is high enough ($\sim$0.1 GK) to ignite the 3$\alpha$-process, creating a small abundance of carbon~\cite{eze71}, e.g., $X_\mathrm{^{12}C}$ $\sim$ 10$^{-9}$ to serve as a catalyst and initiate the CNO cycles. The stellar evolution simulations of Clarkson \& Herwig~\cite{cla21} using the NACRE rate set~\cite{ang99} that supersedes the rates used by Heger \& Woosely~\cite{hw10} confirmed that the CNO cycling takes place at a core H-burning temperature of up to $\sim$0.12~GK. Their nucleosynthesis calculations found that it was unlikely that large amounts of Ca could be produced by hot CNO breakout. Their predicted Ca abundance was between $\sim$0.8 and nearly 2 dex lower than required by observations of the most metal-poor stars. If, however, ratio of the $^{19}$F($p$,\,$\gamma$)$^{20}$Ne and the $^{19}$F($p$,\,$\alpha$)$^{16}$O reaction rates were a factor of $\sim6$ higher than that reported in the NACRE compilation~\cite{ang99}, their models could produce Ca at the level observed in ultra-metal poor stars such as SMSS0313-6708. 

SMSS0313-6708 is an ultra-metal poor (UMP) star that is speculated to be a direct decedent of the first generation of stars in the universe that formed after the Big Bang. The observable composition of an UMP star is a time capsule to the environment before the first galaxies formed – complementing the exiting upcoming observations of the James Webb telescope~\cite{jwst}, which is now aiming to give a first look at the earliest stars and galaxies.

Here, the ($p$,\,$\gamma$)/($p$,\,$\alpha$) rate ratio can provide an invaluable tool to diagnose how the first stars evolved and died, and has far-reaching implications on the stellar modeling. If Ca were produced from such hot hydrogen burning, the Ca produced in the later Si-burning phases can fall back onto a central black hole during the supernova~\cite{cha18}, which is a key ingredient in the prevailing faint supernova with efficient fallback scenario. Otherwise, such a scenario has to be revised, or an alternative source must be validated. Other potential sources include a convective-reactive light Pop III $i$-process~\cite{cla18} or Ca synthesis from explosive burning~\cite{lim12}. Therefore, an accurate determination of the $^{19}$F($p$,\,$\gamma$)$^{20}$Ne rate around 0.1~GK is extremely important to pin down the origin of Ca made by Pop III stars, as well as validating the stellar evolution models.

In the center-of-mass energy region of primary astrophysical interest ($E_\mathrm{c.m.}$ $<1\,$MeV), very limited experimental data are available for the $^{19}$F($p$,\,$\gamma$)$^{20}$Ne reaction due to the very strong 6.130-MeV $\gamma$-ray background from the competing $^{19}$F($p$,\,$\alpha\gamma$)$^{16}$O channel.  This makes measurements of such small $^{19}$F($p$,\,$\gamma$)$^{20}$Ne cross section extremely difficult. Most of the previous experiments detected the $>$11~MeV primary transition to the first excited state of $^{20}$Ne~\cite{sin54,far55,kes62,ber63,sub79} using small-volume NaI(Tl) detectors with low resolution and efficiency. The earlier measurements also suffered from pileup from the 6.130-MeV $\gamma$ rays because of insufficient energy resolution to separate the two components. Later, Couture~\textit{et al.}~\cite{cou08} developed a coincident detection technique (between HPGe and NaI detectors) to measure the $^{19}$F($p$,\,$\gamma$)$^{20}$Ne and $^{19}$F($p$,\,$\alpha\gamma$)$^{16}$O reactions over an energy range of $E_\mathrm{c.m.}$ = 200--760~keV. Due to their limited sensitivity, only an upper-limit for the strength of the $E_\mathrm{c.m.}$ = 213~keV resonance was given and no estimate was made for the 225~keV resonance, although it had been observed as a resonance in the $^{19}$F($p$,\,$\alpha\gamma$)$^{16}$O reaction by Spyrou~\textit{et al.}~\cite{spy00}. Williams~\textit{et al.}~\cite{wil21} measured a factor of 2 larger strength value than that of Couture~\textit{et al.} for the 323-keV resonance by using the inverse kinematics method, because their measurement also included contribution owing to the ground-state transition.
Recently, deBoer \textit{et al.}~\cite{deB21} reanalyzed the available $^{19}$F($p$,\,$\gamma$)$^{20}$Ne and $^{19}$F($p$,\,$\alpha$)$^{16}$O experimental data in the $R$-matrix framework, and estimated the corresponding rates for these two reactions. The Pop III star Ca production problem, however, was even intensified with their estimated ratio of the $^{19}$F($p$,\,$\gamma$)$^{20}$Ne and $^{19}$F($p$,\,$\alpha$)$^{16}$O rates, where the ratio was about a factor of 4 lower than that of NACRE.

To date, there is a scarcity of experimental data in the energy region below $E_\mathrm{c.m.}$ $\approx$ 0.35~MeV. To provide an accurate thermonuclear rate, it is of paramount importance to directly measure the $^{19}$F($p$,\,$\gamma$)$^{20}$Ne reaction cross section in this region.  Since the cosmic-ray background radiation is very strong on the Earth's surface, i.e., above-ground laboratories, direct measurements of such small cross sections are extremely challenging. The China Jinping underground laboratory (CJPL) is located in a traffic tunnel of a hydropower station under Jinping Mountain, in the southwest of China~\cite{kang10} with about 2400\,m of vertical rock overburden.  By this measure, it is the deepest operational underground laboratory for particle and nuclear physics experiments in the world. It offers a great reduction in the muon and neutron fluxes by six and four orders of magnitude, respectively, compared to those at the Earth’s surface.  
The cosmic-ray induced background measured at CJPL~\cite{wu13} is about two orders of magnitude lower than that in LUNA (1400\,m thick dolomite rocks)~\cite{bro10}. With such a unique ultra-low-background environment, the Jinping Underground Nuclear Astrophysics Experiment (JUNA)~\cite{liu16} was initiated, and we have performed a $^{19}$F($p$,\,$\gamma$)$^{20}$Ne direct measurement campaign as one of the Day-one experiments. 

\begin{figure}[!h]
\begin{center}
\includegraphics[width=89mm]{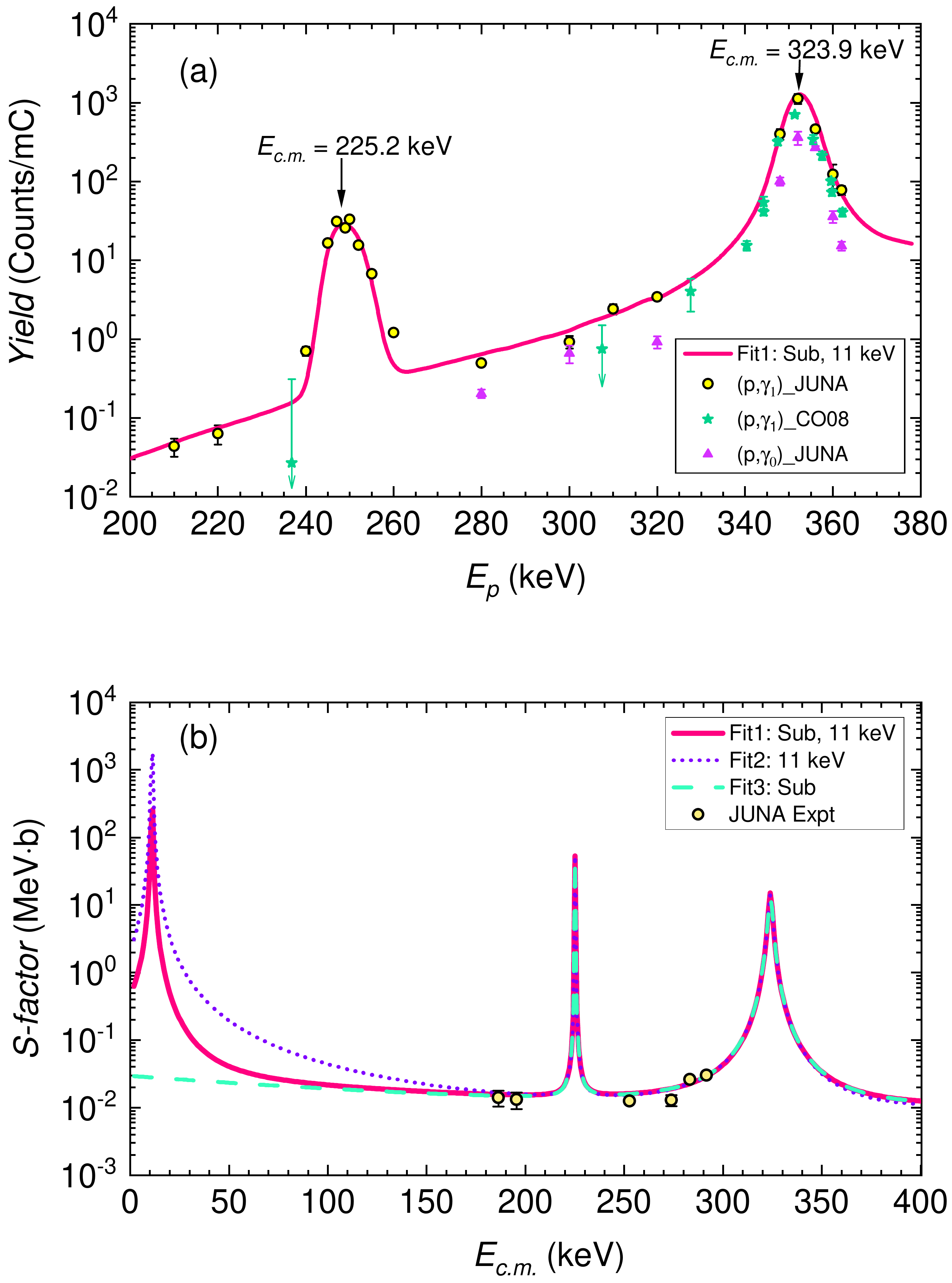}
\caption{\label{fig:exp-yields} \textsl{(a)} Experimental yields of the $^{19}$F($p$,\,$\gamma_{0,1}$)$^{20}$Ne reaction measured at JUNA. Previous experimental ($p$,\,$\gamma_1$) data~\cite{cou08}, which overlap with the present energy regime, are shown for comparison. The \texttt{Geant4} simulated yield curve is depicted using the $R$-matrix fit (``Fit1''). Here, $E_p$ denotes the proton beam energy delivered from the accelerator.
\textsl{(b)} Three probable astrophysical $S$-factor curves for the $^{19}$F($p$,\,$\gamma_1$)$^{20}$Ne reaction fitted by the $R$-matrix calculations. Six data points are derived from the present JUNA experiment. The uncertainties are purely statistical. The error bars are invisible where they are smaller than the data-point size. See Extended Data Fig.~4 for fitting covariance matrix.}
\end{center}
\end{figure}

The experiment was performed in normal kinematics using a high-current 400~kV electrostatic accelerator~\cite{wu16} at CJPL. 
A well-focused, high intensity proton beam uniformly impinged on a $^{19}$F water-cooled target with a current up to $\approx$1~mA. The experimental setup is shown in Extended Data Fig.~1. Durable implanted $^{19}$F targets~\cite{zha21} were used in both $^{19}$F($p$,\,$\alpha\gamma$)$^{16}$O~\cite{PRL} and $^{19}$F($p$,\,$\gamma$)$^{20}$Ne experiments. The $\gamma$ rays were detected using a nearly 4$\pi$ BGO detector array that was also employed in the preceding JUNA experiments~\cite{PRL,SuJ}. The typical $\gamma$-ray spectra are shown in Extended Data Fig.~2.
Owing to the different detection efficiency, the contribution of the summing $\gamma_\mathrm{sum}$ rays (at $\sim13\,$MeV) from $^{19}$F($p$,\,$\gamma$)$^{20}$Ne reaction has been separated into two components: one involves only the transition to the ground state (\textit{g.s.}) in $^{20}$Ne, hereafter referred to as ($p$,\,$\gamma_0$); another involves all transitions through the $1.634\,$MeV first excited state to the \textit{g.s.} in $^{20}$Ne, hereafter referred to as ($p$,\,$\gamma_1$). This way, the ($p$,\,$\gamma_1$) component can be determined precisely based on the coincident technique described here, because the nearby heavy summing signal (at $\sim$12 MeV) induced by the 6.130-MeV $\gamma$ rays interferes with the total counts of the summed $\gamma$ rays, as do those $\gamma$ rays from the $^{11}$B($p$,\,$\gamma$)$^{12}$C contamination reaction at lower proton energies. As shown in the inset of Extended Data Fig.~2, the 1.634$\rightarrow$\textit{g.s.} transition can be clearly observed by gating on the summing $\gamma_\mathrm{sum}$ rays, which correspond to the $^{19}$F($p$,\,$\gamma_1$)$^{20}$Ne component. Figure~\ref{fig:exp-yields}(a) shows the resulting $1.634\,$MeV $\gamma$-ray yields obtained using this coincidence technique.

A new resonance has been discovered at $E_\mathrm{c.m.}=225\,$keV for the first time, well below the well-known resonance at $E_\mathrm{c.m.}=323\,$keV. For the known $323\,$keV resonance, the $\gamma$-yield ratios between the ($p$,\,$\alpha\gamma$) and ($p$,\,$\gamma$) channels obtained are shown in Extended Data Fig.~3.  We determined partial strengths of $\omega\gamma_{(p,\gamma_1)}=2.09\pm0.21\,$meV and $\omega\gamma_{(p,\gamma_0)}=1.07\pm0.21\,$meV, respectively. 
Thus, its total strength is determined to be $\omega\gamma_{(p,\gamma_\mathrm{tot})}=3.16\pm0.33$~meV, where the statistical and systematical errors are 0.23~meV and 0.24~meV, respectively. The present $\omega\gamma_{(p,\,\gamma_1)}$ value is about a factor of 1.5 larger than the previous value of $1.38\pm0.44\,$meV~\cite{cou08}. Both values agree within a 2-$\sigma$ uncertainty, but our value has much improved precision. 
In addition, our total value of $\omega\gamma_{(p,\,\gamma_\mathrm{tot})}$ is consistent with the NACRE adopted value of $5\pm3\,$meV, as well as with the recently reported value of $3.3^{+1.1}_{-0.9}\,$meV ~\cite{wil21} that was also sensitive to the direct capture to g.s., but with 4--7 times better precision. 
For the newly observed $225\,$keV resonance, its strength is determined to be $\omega\gamma_{(p,\,\gamma_1)}$ = ($4.19\pm0.33$)$\times10^{-2}\,$meV based on the yield ratio between the $323\,$keV ($p$,\,$\alpha\gamma$) and $225\,$keV ($p$,\,$\gamma_1$) resonances. Our resonance strengths were all determined relatively to the well-known ($p$,\,$\alpha\gamma$) strength of the $E_\mathrm{c.m.}$ = 323 keV resonance.
Here, the yield corresponds to the integrated $\gamma$-ray counts (corrected for efficiency) under the yield curve over the resonance.  We find that the ($p$,\,$\gamma_0$) contribution is negligibly small in the energy region below  $E_\mathrm{c.m.}\approx322\,$keV (see Figure~\ref{fig:exp-yields}(a)), and hence $\omega\gamma_{(p,\,\gamma_\mathrm{tot})}\approx\omega\gamma_{(p,\,\gamma_1)}$ for this resonance.
For the previously theorized $E_\mathrm{c.m.}=213\,$keV resonance, estimates placed upper limits on the strength at $1.3\times10^{-3}\,$meV~\cite{ang99} and $9.3\times10^{-4}\,$meV~\cite{cou08}; we now firmly constrain its strength to be $<4.2\times10^{-3}\,$meV (i.e., less than $10\%$ of that of the $225\,$keV resonance) based on the present experimental data.  
Table~\ref{tab1} summarizes the resonance properties.

\begin{table*}
\caption{Relevant resonance strengths $\omega\gamma_\mathrm{tot}$ determined for the $^{19}$F($p$,\,$\gamma$)$^{20}$Ne reaction, with total errors listed in the parentheses.
$R$-matrix fit parameters are tabulated, including a sub-threshold and near-threshold $11\,$keV resonances as shown in Figure~\ref{fig:exp-yields}(b). The fit includes the additional levels and from Ref.~\cite{deB21} as fixed background terms. See Method for details.  
\label{tab1}}
%a = Kious (thesis)
%b = deBoer (PRC)
%c = Spyrou (PRC)
\footnotesize
\begin{tabular}{cccccccccc}
\hline \hline
(keV) & (MeV) & & \multicolumn{2}{c}{$\omega\gamma_\mathrm{tot}$ (meV)} & & \multicolumn{1}{c}{(fm$^{-1/2}$)} & & \multicolumn{2}{c}{(eV)} \\
\cline{4-5} \cline{7-7} \cline{9-10}
$E_\mathrm{c.m.}$ & $E_x$ & $J^\pi$ & Present & NACRE~\cite{ang99} & & ANC & & $\Gamma_{\alpha_2}$ & $\Gamma_{\gamma_1}$ \\
\hline
$-448$ & 12.396(4)$^a$ & 1$^+$ &    &  & & 15$^b$ & & 60$^{+40}_{-30}$ & $<$3.4 \\
11 & 12.855(4)$^a$ & 1$^+$  & &  & & 1.14$\times$10$^{-28}$~$^c$ & & $-590^{+230}_{-290}$ & $<$4.8 \\
212.7(10) & 13.057 & 2$^-$  & $<$4.2$\times$10$^{-3}$ &  $<$1.3(13)$\times$10$^{-3}$ & & & & & \\
225.2(10) & 13.069 & 3$^-$  & 4.19(33)$\times$10$^{-2}$ &  & & & & & \\
323.9 & 13.168(2)$^b$ & 1$^+$ & 3.16(33) & 5(3) & & & & & \\
\hline \hline
\end{tabular}
\end{table*}

A multilevel, multichannel $R$-matrix analysis, using the code \texttt{AZURE2}~\cite{azu10,ube15}, was used to fit the data. 
The $R$-matrix analysis is an extension of that presented in Ref.~\cite{deB21}, and includes all those data together with the new CJPL ($p$,\,$\gamma_1$) and ($p$,\,$\alpha\gamma$) data. 
With this method, various possible contributions can be strictly constrained. The curve shown in Figure~\ref{fig:exp-yields}(a) represents the \texttt{Geant4}~\cite{geant} simulated results by using one of the lowest $\chi^{2}$ $R$-matrix fits (``Fit1'') to the $S$-factor data. Figure~\ref{fig:exp-yields}(b) shows six off-resonance data points derived from the present JUNA experiment. 
Numeric samples of the $S$-factors and the associated uncertainties in the off-resonance region are tabulated in Extended Data Table~1. 
We present the three best $R$-matrix fits. Here, ``Sub" denotes the 1$^+$ sub-threshold state at $E_x=12.396$~MeV, and ``$11\,$keV''denotes the $11\,$keV 1$^+$ resonance at $E_x=12.855\,$MeV. For example, the label ``Fit1: Sub, $11\,$keV'' indicates the $R$-matrix fit including both the sub-threshold state and the $11\,$keV resonance. The nuclear level properties in the $R$-matrix fits were co-varied over a large parameter space. 
The resultant resonance properties deduced from the $R$-matrix fits are listed in Table~\ref{tab1} (See Methods for details of the $R$-matrix calculations).

\begin{figure}[!ht]
\begin{center}
\includegraphics[width=150mm]{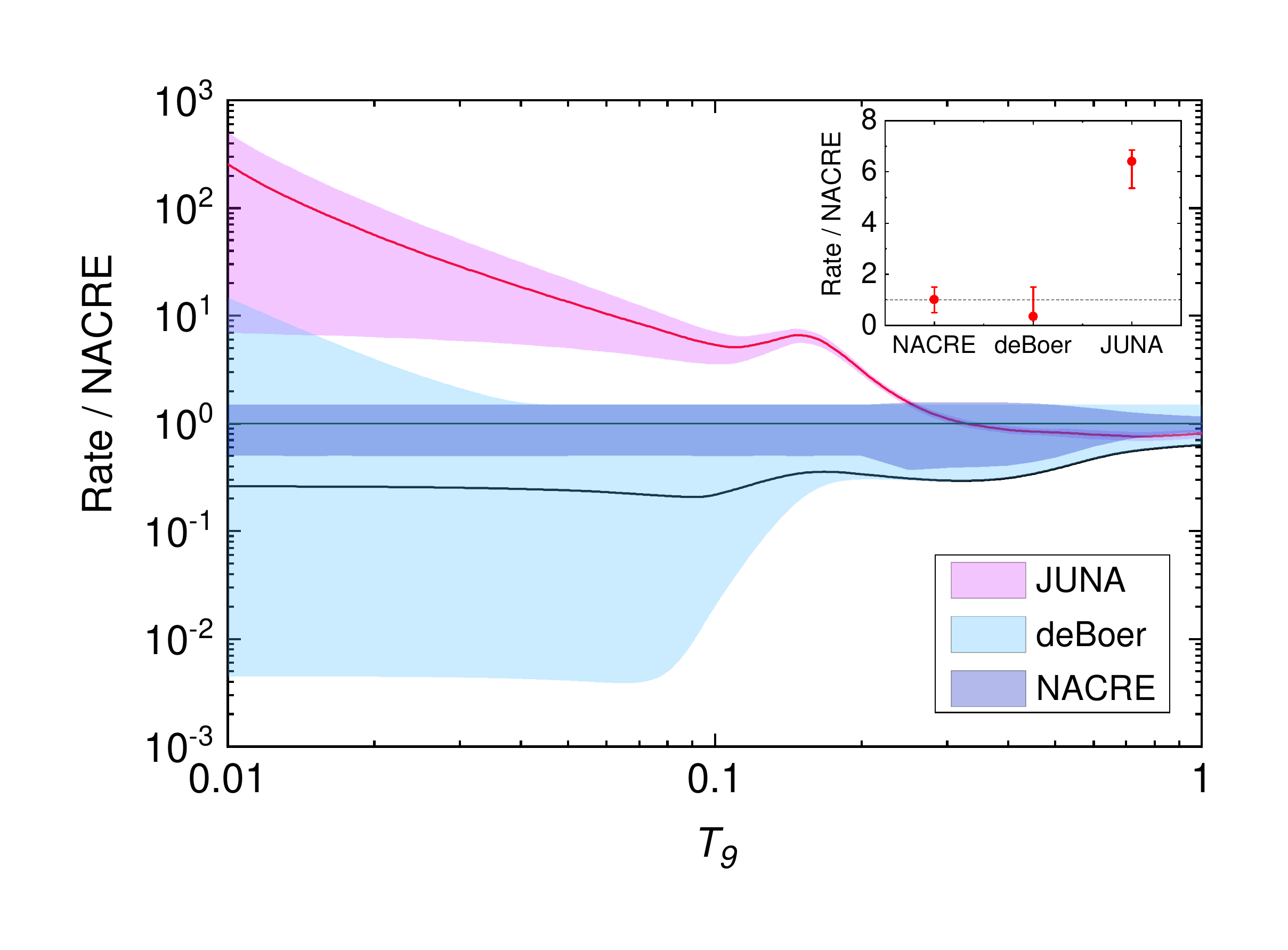}
\vspace{-8mm}
\caption{\label{fig:ratio-lit} Ratio of the present (labelled as JUNA) relative to NACRE's rate~\cite{ang99}.  The corresponding ratio for deBoer \textit{et al.\/}'s rate~\cite{deB21} is also shown for comparison. The associated uncertainties are shown as the colored bands. The \textsl{inset} shows the ratios at a temperature of $0.14\,$GK.}
\end{center}
\end{figure}

The thermonuclear $^{19}$F($p$,\,$\gamma$)$^{20}$Ne reaction rate as a function of temperature is calculated by numerical integration of the $S$-factors shown in Figure~\ref{fig:exp-yields}(b)~\cite{rol88}. The mean rate and the associated uncertainties (Low and High limits) are obtained in a temperature region of 0.01--1~GK and presented in Extended Data Table~2. The ratios between the present rate and the NACRE recommended rate are shown in Figure~\ref{fig:ratio-lit}. It shows that our new rate is enhanced by a factor of $5.4$--$7.4$ at the temperature around $0.1\,$GK. This enhancement is attributed to the newly observed $225\,$keV resonance. In addition, our new rate is about $200$ times larger at temperatures around $0.01\,$GK mainly because of the $11\,$keV resonance~\cite{PRL}. The uncertainty of the present rate is drawn as a colored band, which we estimate based on the uncertainties of the resonance strengths and $R$-matrix calculations. The uncertainties in the present $S$-factor and rate over the range of astrophysical interest have been significantly reduced compared to previous estimates~\cite{deB21}. 

deBoer \textit{et al.}~\cite{deB21} recommended a $^{19}$F($p$,\,$\alpha$)$^{16}$O mean rate quite similar to that of NACRE~\cite{ang99}. Thus, we adopt NACRE's $^{19}$F($p$,\,$\alpha$)$^{16}$O rate as our reference, and hence obtain an enhancement factor of $5.4$--$7.4$ for the ($p$,\,$\gamma$)/($p$,\,$\alpha$) rate ratio relative to that of NACRE~\cite{ang99} at around $0.1\,$GK. We find an even larger enhancement below $\sim0.08\,$GK. By a simple scaling argument to the model calculations in Refs.~\cite{cla21,deB21}, the observed Ca abundances in the oldest known SMSS0313-6708 star can now be reproduced reasonably with our new $^{19}$F($p$,\,$\gamma$)$^{20}$Ne rate. 

\begin{figure*}[!ht]
\begin{center}
\includegraphics[width=0.59\textwidth]{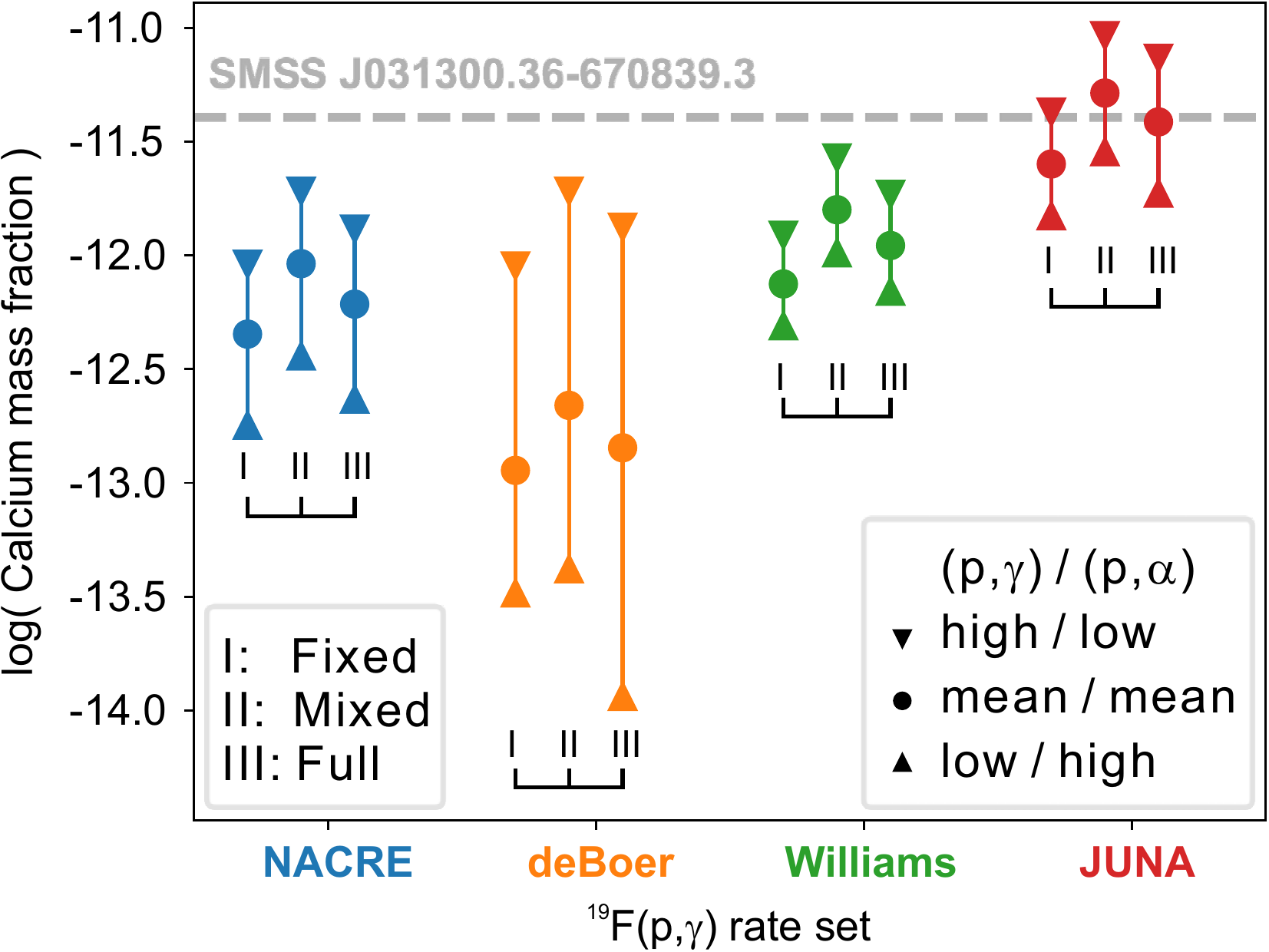}
\hfill
\includegraphics[width=0.4\columnwidth,viewport=420 0 770 400,clip=true]{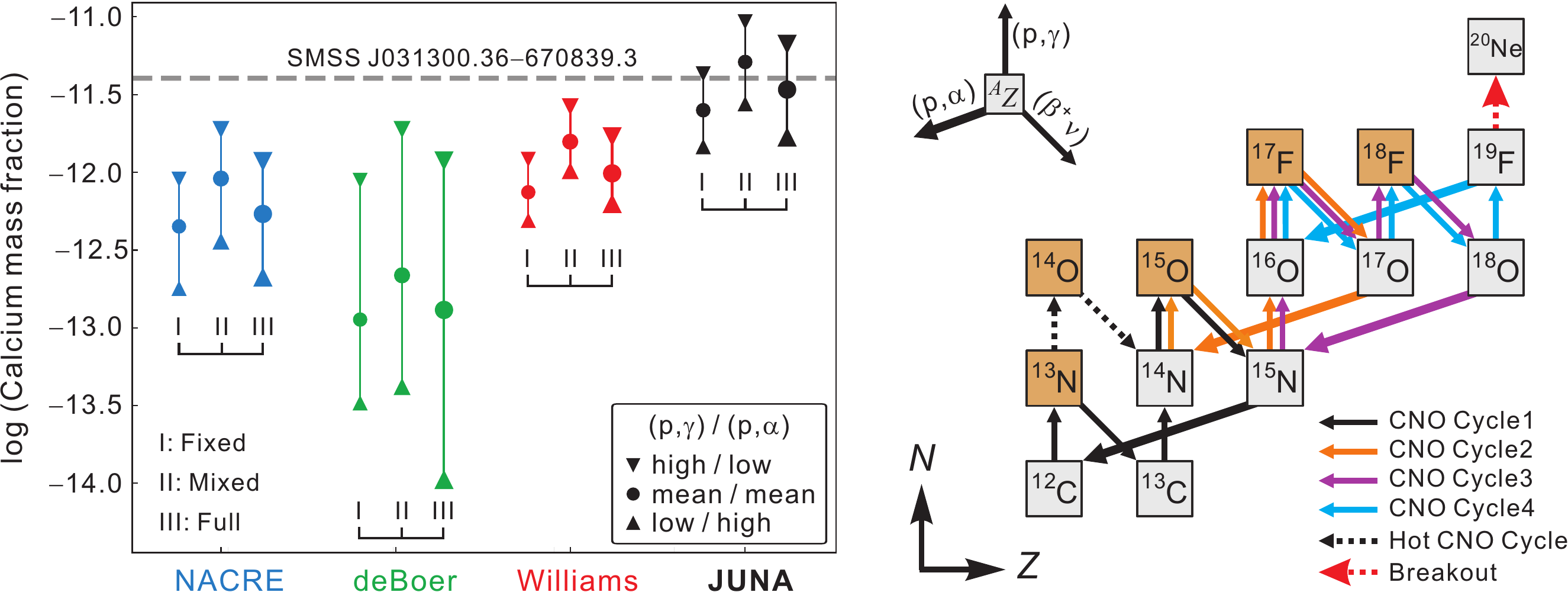}
%\includegraphics[width=0.6\textwidth]{Ca_reaclib_all.pdf}
%\\\vskip -6\baselineskip
%\includegraphics[width=0.6\columnwidth,viewport=420 0 770 400,clip=true]{Figure3.pdf}
\caption{
\textsl{(Left Panel)}
Range of results for different rate combinations and different modelling techniques.
(\textsf{I}) is for trajectories of fixed temperature and density; (\textsf{II}) is for time-dependent trajectories (Figure~S2) that include the effect of mixing due to convection; (\textsf{III}) is for yields from full stellar models.  \label{fig:YieldAll} 
\textsl{(Right Panel)} The four classical CNO cycles~\cite{wie99} (\textsl{solid lines}) and the hot CNO cycle shortcut (\textsl{black dotted lines}).  The breakout $^{19}$F($p$,~$\gamma$)$^{20}$Ne reaction route is indicated as \textsl{red dotted arrow}. See Supplemental Information for more details.
\label{fig:network} 
}
\end{center}
\end{figure*}

\begin{figure*}[!ht]
\begin{center}
\includegraphics[width=89mm]{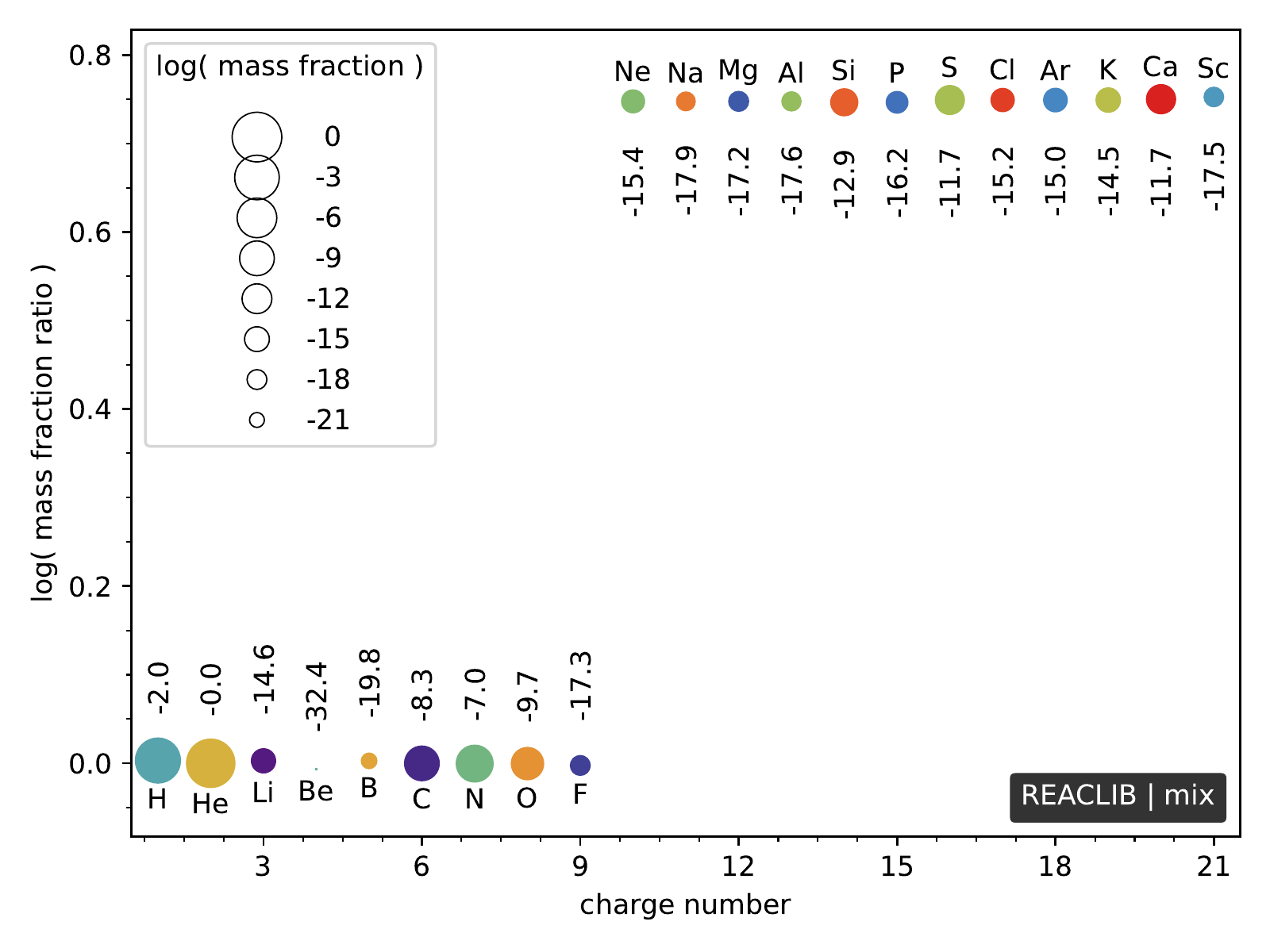}
\end{center}
\caption{Ratio of final abundances of using our JUNA mean $^{19}$F($p$,\,$\gamma$) rate compared to using the NACRE~\cite{ang99} mean rate. Both use the NACRE mean $^{19}$F($p$,\,$\alpha$) rate. The ratio is shown at core hydrogen depletion, at a hydrogen mass fraction of $0.01$ and using the mixing model. The linear size of the symbol indicates the logarithm of the absolute mass fraction (see \textsl{legend}).  Their numerical values are given above/below the symbols.}
\end{figure*}

We have investigated the impact of the thermonuclear $^{19}$F($p$,\,$\gamma$)$^{20}$Ne rate on a range of nucleosynthesis modelling techniques, and the calcium production is summarized in Figure~\ref{fig:YieldAll} (\textsl{Left Panel}). Our studies comprise simple trajectories (see Extended Data Table~3), new mix models (see Extended Data Table~4), and full stellar models (see Extended Data Table~5) calculations. We find that all our nucleosynthesis models can reproduce the observed calcium production. We conclude that the $^{40\!}$Ca observed in the oldest known ultra-iron poor stars (e.g., SMSS0313-6708) may indeed originate in hydrostatic burning in Pop III stars, requiring only the supernova ejection of their outer layers, whereas the metal-rich core may collapse to a black hole. Previously, the ejection of the metal-rich core was required as the source of calcium abundance observed in the oldest stars.  On the contrary, here we show a much stronger breakout from a ``warm" CNO cycle scenario via $^{19}$F($p$,\,$\gamma$)$^{20}$Ne, which significantly increases the production of Ne -- Ca. Figure~4 shows the ratio of final abundances of using our JUNA mean $^{19}$F($p$,\,$\gamma$) rate compared to using the NACRE mean rate. The production of all elements beyond $Z=9$ is shifted by a constant factor and hence can be well represented by single species, e.g., the double-magic nucleus $^{40}$Ca that was observed in ultra-metal poor stars. This clearly shows the bottle-neck nature of the $^{19}$F($p$,\,$\gamma$)---$^{19}$F($p$,\,$\alpha$) branching point. See Supplemental Information for more details.

To conclude, we have directly measured the $^{19}$F($p$,\,$\gamma$)$^{20}$Ne reaction down to the unprecedentedly low energy point of $E_\mathrm{c.m.}$ $\approx186$~keV by exploiting the extremely low background environment deep underground, high-intensity beam and newly developed durable target(s). All these unique and featured conditions allowed us to measure this crucial reaction at the stellar energy region, which is inaccessible in the above-ground laboratories for decades. We have discovered a new key resonance and determined for the first time a precise thermonuclear rate over the temperature region of astrophysical importance. Our enhanced new rate leads to a stronger breakout in a ``warm" CNO scenario as the origin of the calcium discovered in the oldest, ultra-iron poor stars. Our results provide a strong experimental foundation to the faint supernova model of first-generation primordial stars as source for the observed chemical abundance signature. The astrophysical implications of our new rate on novae, X-ray bursts, AGB stars, and other star sites are still subject to future detailed investigations.

\vspace{1cm}
\noindent
\textbf{Acknowledgments}
We thank the staff of the CJPL and Yalong River Hydropower Development Company (N.C. Qi, W.L. Sun, X.Y. Guo, P. Zhang, Y.H. Chen, Y. Zhou, J.F. Zhou, J.R. He, C.S. Shang, M.C. Li) for logistics support. We thank F. Herwig, Y. Sun and S.E. Malek for discussions.
We acknowledge support from the National Natural Science Foundation of China (Nos. 11825504, 11490560, 12075027, 12125509). R.D. and M.W. were supported by the NSF through Grant No. Phys-2011890.  R.D., M.W., and A.H. were supported by the Joint Institute for Nuclear Astrophysics through Grant No. PHY-1430152 (JINA Center for the Evolution of the Elements). 
A.H. was supported by the Australian Research Council (ARC) Centre of Excellence (CoE) for Gravitational Wave Discovery (OzGrav) through project number CE170100004, by the ARC CoE for All Sky Astrophysics in 3 Dimensions (ASTRO 3D) through project number CE170100013.
D.K. acknowledges the support of the Romanian Ministry of Research and Innovation under research contract 10N/PN 19 06 01 05.
\\
\textbf{Author contributions: }
M.W. proposed the original idea of this research.
J.H. and W.L. proposed this JUNA experiment. L.Z., J.H. designed the experimental setup and led all the tests and experiments, and performed the data reduction and analysis.
R.D., and M.W. performed the $R$-matrix analysis.
A.H. made the astrophysical model calculation and interpretation.
J.S., Y.C., X.L., H.Z., X.J., L.W., Z.L., and L.S. participated in the experiment.
J.H., A.H., D.K., R.D., M.W., W.L. prepared the draft of the manuscript.
D.K. made major contributions to the manuscript polishing.
All authors read the manuscript, gave comments, suggested changes, and agreed with the final version.
L.Z., F.C., Y.C., and Z.Z. took main responsibility for the operation of the JUNA accelerator.
J.S., and Z.L. developed the 4$\pi$ BGO detector array, and J.W. developed the DAQ system. L.S., Q.W., J.L., and H.Z. designed and constructed the ECR ion source. B.C., L.C., R.M., and G.L. designed and constructed the 400-kV accelerator.
J.H. supervised the experiment and verified that the data were acquired correctly as a P.I. of this sub-project.
W.L. leads the JUNA project, and J.C. leads the CJPL.
\\
\textbf{Competing interests: }The authors declare no competing interests.
\\
\textbf{Data availability}
Experimental data taken at JUNA are proprietary to the collaboration
but can be made available from the corresponding authors upon reasonable request. 
%Source data are provided with this paper.
\\
\textbf{Code availability}
The $R$-matrix code can be made available
upon request to R.D. (e-mail: richard.james.deboer@gmail.com).
\\
%\textbf{Additional information}
%\\
\textbf{Extended data} is available for this paper at \textbf{https://doi.org/10.1038/s41586-022-05230-x}
\\
\textbf{Correspondence and requests for materials} should be addressed to Jianjun He, Michael Wiescher or Weiping Liu.

\bibliography{scibib}

\bibliographystyle{Science}

% For your review copy (i.e., the file you initially send in for
% evaluation), you can use the {figure} environment and the
% \includegraphics command to stream your figures into the text, placing
% all figures at the end.  For the final, revised manuscript for
% acceptance and production, however, PostScript or other graphics
% should not be streamed into your compliled file.  Instead, set
% captions as simple paragraphs (with a \noindent tag), setting them
% off from the rest of the text with a \clearpage as shown  below, and
% submit figures as separate files according to the Art Department's
% instructions.

%\clearpage
%
%\noindent {\bf Fig. 1.} Please do not use figure environments to set
%up your figures in the final (post-peer-review) draft, do not include graphics in your
%source code, and do not cite figures in the text using \LaTeX\
%\verb+\ref+ commands.  Instead, simply refer to the figure numbers in
%the text per \textit{Science\/} style, and include the list of captions at
%the end of the document, coded as ordinary paragraphs as shown in the
%\texttt{scifile.tex} template file.  Your actual figure files should
%be submitted separately.

\section*{Methods}

\noindent
\smallskip
\textbf{JUNA experiment.}
The Jinping Underground Nuclear Astrophysics Experiment (JUNA)~\cite{liu16} was initiated in 2015. One of the
Day-one goals~\cite{he16} was to directly measure the $^{19}$F($p$,\,$\alpha\gamma$)$^{16}$O reaction at Gamow energies. The measurement was accomplished
and results were published elsewhere~\cite{PRL}. The present $^{19}$F($p$,~$\gamma$)$^{20}$Ne experiment was immediately followed that ($p$,\,$\alpha\gamma$) run with the same experiment setup, acting as one of the Day-one campaigns. In combination with the ultra-low background environment, the strong beam intensity, the durable target, as well as the coincidence technique, it ultimately makes this direct $^{19}$F($p$,~$\gamma$)$^{20}$Ne measurement possible.

The schematic view of the experimental setup is shown in Extended Data Fig.~1. A proton beam from the accelerator was undulated over a rectangular area of about 4$\times$4~cm$^2$ by oscillating the magnetic field of the beam deflector. A well-focused, intense beam was uniformly spread across the target, mitigating damage to the target. The scanning proton beam was collimated by two apertures ($\phi$15 upstream and $\phi$12~mm downstream) and then impinged on a water-cooled target, where the beam current reached up to 1~mA, with a spot size of about $\phi$10~mm. An inline Cu shroud cooled to LN$_{2}$ temperature extended close to the target to minimize carbon build-up on the target surface. Together with the target, the Cu shroud constituted the Faraday cup for beam integration. A negative voltage of 300~V was applied to the shroud to suppress secondary electrons from the target. A very strong and durable implanted $^{19}$F target~\cite{zha21} was utilized in this work. The optimum scheme for target production is: first, implanting $^{19}$F ions into the pure Fe backings with an implantation energy of 40 keV, and then sputtering a 50-nm thick Cr layer to further prevent the fluorine material loss. The 4$\pi$ Bi$_4$Ge$_3$O$_{12}$ (BGO) detector array specially designed for the JUNA project is composed of eight identical segments with a length of 250~mm and a radial thickness of 63~mm, each covering a 45$^\circ$ azimuthal angle. For the 6.130-MeV $\gamma$ rays, the total absolute detection efficiency was $\approx$58\%, with a $\approx$6\% energy resolution achieved by alcohol--cooling the BGO crystals ($\approx$--5$^\circ$C). To further suppress the natural background emitted from the rocks and induced by neutron capture reactions, the BGO array was passively shielded by 5-mm copper, 100-mm lead and 1-mm cadmium, respectively. By adjusting the beam intensity in each run, the counting rate of the BGO array was limited to about 10~kHz to prevent the signal pileups and reduce the dead-time of the DAQ system. In addition, the waveforms of pulses were recorded in the DAQ system to monitor the pileup events during the experiment. We found that the pileup events are very rare, and can be completely ignored.

Extended Data Fig.~2 shows the typical $\gamma$-ray spectra taken for two typical energy points, (a) at $E_{p}$ = 356~keV and (b) at $E_{p}$ = 250~keV.
Here, $E_p$ denotes the proton beam energy delivered from the accelerator, and the real bombarding energy on the fluorine atoms is reconstructed by taking into account the energy loss through the Cr protective layer with a \texttt{Geant4} simulation~\cite{geant}.
It shows that the 6.130-MeV $\gamma$ rays (from the $\alpha\gamma_2$ channel) dominate the whole spectra, while the 6.917-MeV (from the $\alpha\gamma_3$ channel) and 7.117-MeV (from the $\alpha\gamma_4$ channel) $\gamma$ rays observed at certain proton energies, only make a maximum contribution of $\approx$2.4\% in the energy region studied in this work. Here, we are mainly concerned with the summing $\gamma$-ray peak for the targeted $^{19}$F($p$,~$\gamma$)$^{20}$Ne channel around 13 MeV. The $\gamma$ rays induced by the $^{11}$B, $^{12}$C and $^{13}$C contaminants were observed at certain energies, and their origins were clearly identified~\cite{zha21,PRL}.
The $^{19}$F target material loss was monitored and found to be negligible since the total beam dose utilized in this measurement was only about 41~C, which was consistent with prior expectations~\cite{zha21}.

A precise determination of the absolute $^{19}$F number density is challenging because of the complicated target structure and the unknown self-sputtering rate during the implantation procedure.
Similar to previous work~\cite{PRL}, we derived the ($p$,\,$\gamma$) strengths of the $225\,$keV and $323\,$keV resonances relative to the well-known ($p$,\,$\alpha\gamma$) strength of the 323-keV resonance.  Its strength was evaluated as $\omega\gamma_{(p,\,\alpha\gamma)}=23.1\pm0.9\,$eV in NACRE, i.e., with an uncertainty of about 4\%. For the 323-keV resonance, the ratios between ($p$,\,$\alpha\gamma$) and ($p$,\,$\gamma$) yields are obtained at five energy points over the resonance, by comparing the corresponding $\gamma$-ray counts corrected by the efficiency.
The corresponding ratios are shown in Extended Data Fig.~3.
The weighted average ratios and the associated uncertainties are plotted as the solid and dashed lines, respectively.  We find weighted average ratios of ($p$,\,$\alpha\gamma$)/($p$,\,$\gamma_1$) and ($p$,\,$\alpha\gamma$)/($p$, \,$\gamma_0$) of (1.11$\pm$0.07)$\times$10$^{4}$ and (2.15$\pm$0.39)$\times$10$^{4}$, respectively.

\smallskip
\noindent
\textbf{Astrophysical $S$ factors.}
Selected astrophysical $S$ factors derived for the $^{19}$F($p$,\,$\gamma$)$^{20}$Ne reaction in the non-resonance region are listed in Extended Data Table~1, which are shown in Figure~1(b) (with statistical uncertainties shown only).
Here, the statistical uncertainties range from 8.3\% to 27.4\% as listed in the last column of Extended Data Table~1.
The systematic uncertainties mainly include the following contributions: 1) a 5\% uncertainty estimated for the \texttt{Geant4} simulation by assuming a 0.5~keV uncertainty in the reconstructed $E_\mathrm{c.m.}$ energy; 2) a 3.9\% uncertainty of the 323-keV resonance strength (from the normalization); and 3) a 5--10\% uncertainty of the 1634-keV $\gamma$-ray coincidence efficiency.  From this, conservatively, we estimate an overall systematic uncertainty of 12\%.

\smallskip
\noindent
\textbf{$R$-matrix fit.}
The temperature relevant to the Population III stars is about 0.1-0.12~GK, corresponding to an energy range around 100 keV.  At such low energies, the Coulomb repulsion between the two interacting particles --  proton and $^{19}$F -- makes the cross sections so small that their laboratory measurement is very challenging due to the low event rate.  Therefore, measurements are typically made at higher energies, and then a model with underlying physical motivation is used to extrapolate to the low energies of interest.  In low energy nuclear physics, $R$-matrix analysis is one of the most successful of these phenomenological reaction models.  The model is both very flexible, applicable to a wide verity of different reactions, yet still has fundamental physical constraints.

At JUNA, we have obtained both $^{19}$F($p$,\,$\gamma$)$^{20}$Ne and $^{19}$F($p$,\,$\alpha\gamma$)$^{16}$O cross section data.  The latter was already described elsewhere~\cite{PRL}.  Whereas the new $^{19}$F($p$,\,$\gamma$)$^{20}$Ne cross section measurements extend to lower energies than any previous measurement, they are still higher in energy than the energy range of interest to astrophysics.  Therefore the \texttt{AZURE2}~\cite{azu10,ube15} $R$-matrix code has been used to simultaneously fit both reactions using our new data.  This $R$-matrix analysis is an extension of earlier work presented in Ref.~\cite{deB21}.

It still remains unknown which resonance contributions dominate at very low energies, thus several $R$-matrix fits were attempted, taking into account different contributions from either a subthreshold resonance (``Sub") or a near-threshold resonance at 11~keV (``11 keV"). The three most probable fit solutions are shown in Figure~1(b). To quantify the uncertainty stemming from the experimental data and the ambiguity in the low energy resonance structure, a Bayesian uncertainty analysis has been performed~\cite{for13}.
Extended Data Fig.~4 shows the covariance matrix from an MCMC analysis that includes $\Gamma_{\gamma_1}$ for both the sub-threshold and near-threshold resonances. The data indicate that at least one of these components is needed. When both are included, the MCMC analysis indicates a non-zero contribution from the sub-threshold resonance contribution and a value that is consistent with zero for the near-threshold resonance. This analysis quantifies an upper limit for the $S$-factor extrapolation as indicated in Figure~1(b) at the 68\% level. A lower limit is calculated using an analysis that only includes the sub-threshold resonance, resulting in a nearly constant low energy $S$-factor. In main Table 1, $E_x$ values fixed to those determined in previous analyses are indicated by $^a$\cite{kio90}, $^b$\cite{deB21} and $^c$\cite{spy00}, where the corresponding uncertainties are adopted. While for the present results, an assumed uncertainty of 1.0 keV are quoted on the resonance energies. The sign of a partial width indicates the sign of the corresponding reduced width amplitude. No uncertainties are quoted for the ANCs of both the bound and near threshold states as none are available from previous literature. Further, the present data only constrain the product of the ANC and the $\gamma$-width and we have chosen to indicate this uncertainty on the $\gamma$-width. The details will be published elsewhere.

\smallskip
\noindent
\textbf{Reaction Rates.}
At each temperature point in Figure~2, three reaction rates were calculated based on the three $S$-factor curves shown in Figure~1(b). The maximum and minimum of the three rates were adopted as the high and low limits, and the average of the maximum and minimum was adopted as the recommended median rate.  Where the rate errors (i.e., low and high limits) are smaller than those caused by the JUNA $S$-factor errors, the $S$-factor errors were adopted accordingly as the total reaction rate error (i.e., low and high limits).  In this way, the present median rate and the associated uncertainties are obtained in a temperature region of 0.01--1 GK. Beyond 1 GK, the NACRE rates can be used.  Extended Data Table~2 lists the presently recommended thermonuclear $^{19}$F($p$, $\gamma$)$^{20}$Ne rates, and the associated uncertainties (low and high values).

The present mean rate can be parameterized by the standard format of~\cite{rau00},
%\begin{widetext}
\begin{eqnarray*}
N_A\langle\sigma v\rangle &=& \mathrm{exp}(-8.41786-\frac{3.8921}{T_9}+\frac{30.2621}{T_9^{1/3}}-66.1213T_9^{1/3}+128.424T_9-94.5477T_9^{5/3}\nonumber \\
&+& 8.01847\ln{T_9})\nonumber \\
&+& \mathrm{exp}(6.20324-\frac{2.65022}{T_9}-\frac{5.03462}{T_9^{1/3}}+6.9107T_9^{1/3}- 0.999798T_9+0.0523095T_9^{5/3}
\nonumber \\
&-& 0.733785\ln{T_9})\nonumber \\
&+& \mathrm{exp}(27.5327-\frac{10.7563}{T_9}-\frac{11.5633}{T_9^{1/3}}-6.36271T_9^{1/3}
+6.62094T_9-1.30661T_9^{5/3}\nonumber \\
&-& 9.70975\ln{T_9})\nonumber \\
&+& \mathrm{exp}(17.4989+\frac{0.0127512}{T_9}-\frac{17.7464}{T_9^{1/3}}-0.540527T_9^{1/3}+2.50115T_9-0.623077T_9^{5/3} \nonumber \\
&-& 1.33705\ln{T_9}) \,
\label{eq5}
\end{eqnarray*}
%\end{widetext}
with a fitting error of less than 1\% over the temperature region of 0.01--1~GK.

\smallskip
\noindent
\textbf{Astrophysical calculations.}
We have investigated the impact of thermonuclear $^{19}$F($p$,~$\gamma$)$^{20}$Ne rate on a range of nucleosynthesis modelling techniques. We have performed a range of full stellar model calculations for a $40\,\mathrm{M}_\odot$ star of initially primordial composition with the \textsc{Kepler} code~\cite{wzw78}. The calcium production is briefly summarized in Figure~3 (Left Panel), and the numerical values are listed in Extended Data Tables~3--5 for simple trajectories, new mix models, and for full stellar model calculations, respectively. 
Please see Supplement Information for more details.

\clearpage
%Extended section
\begin{figure}[!ht]
\begin{center}
\includegraphics[width=183mm]{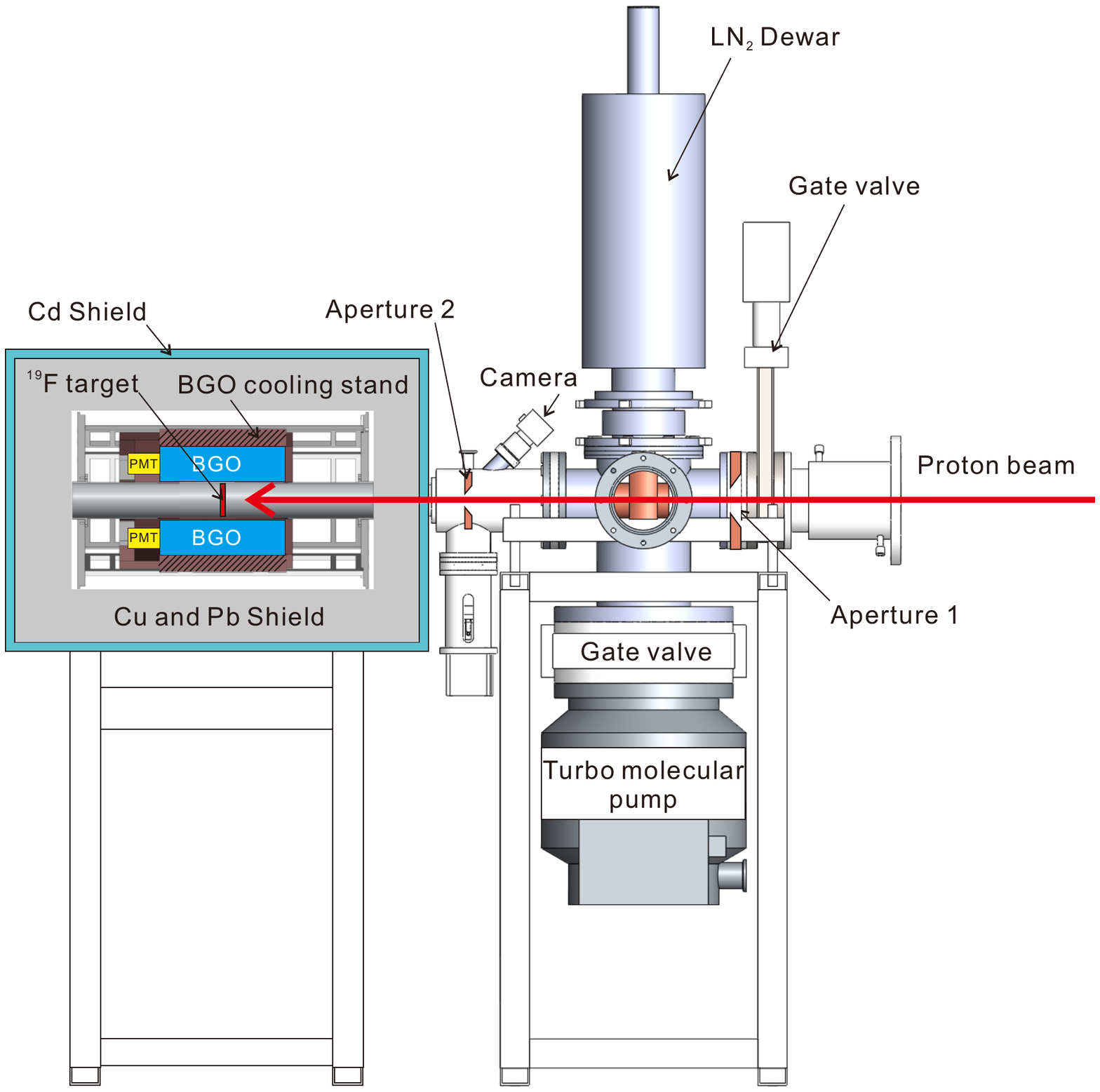}
\vspace{3mm}
\textbf{Extended Data Fig.~1: The schematic view of the experimental setup.} 
\end{center}
\end{figure}

\begin{figure*}[tpb]
\begin{center}
\includegraphics[width=183mm]{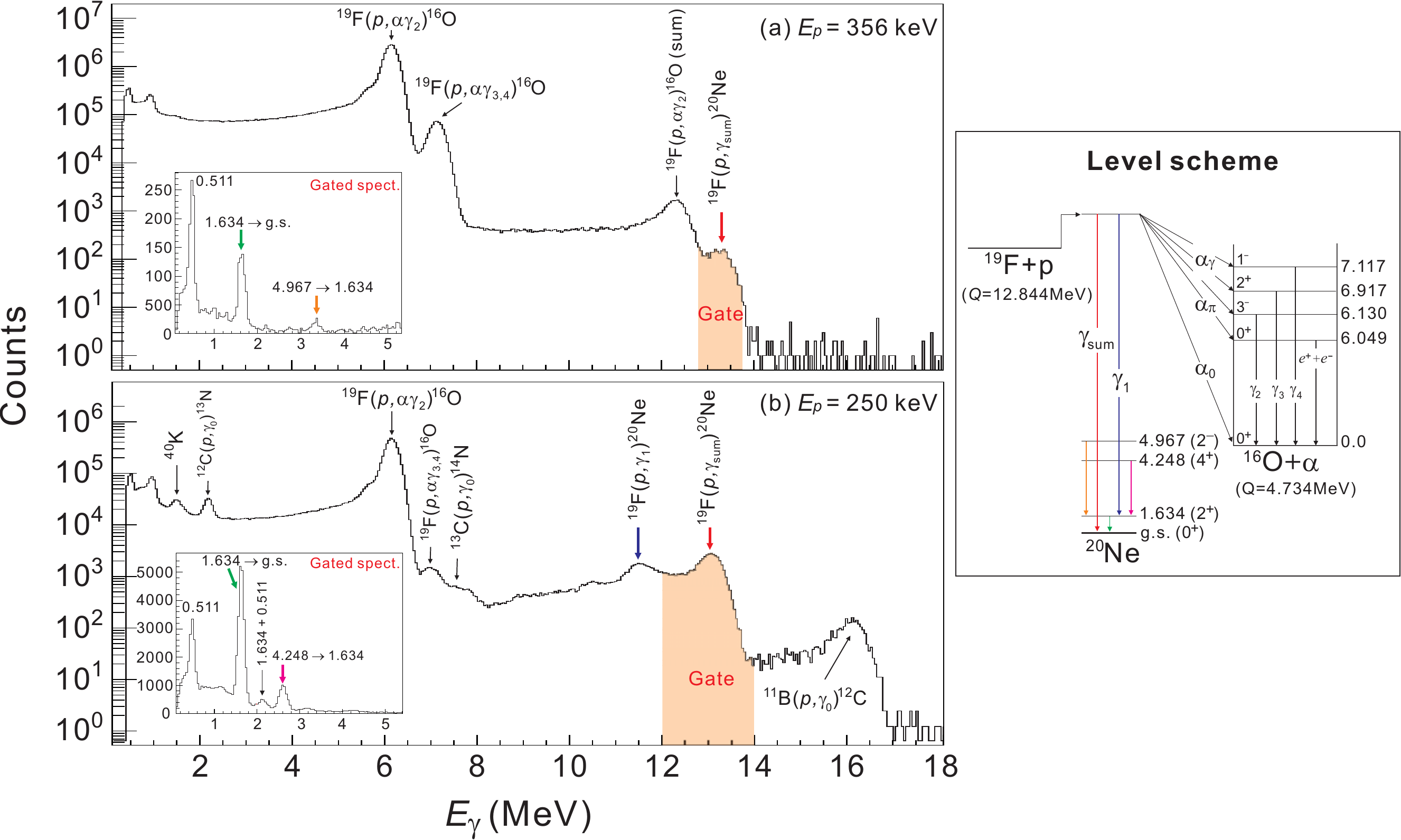}
\end{center}
\textbf{Extended Data Fig.~2: \textsl{(Left Panel)} Typical $\gamma$-ray spectra taken with a 4$\pi$ BGO array at JUNA during proton bombardment of an implanted $^{19}$F target, at proton energies of
(a) 356~keV and (b) 250~keV.} The heavy $\gamma$-ray background from the competing $^{19}$F($p$,\,$\alpha\gamma$)$^{16}$O channel and their summing signals are indicated. The summing $\gamma$-ray peak for the target $^{19}$F($p$,\,$\gamma$)$^{20}$Ne channel is indicated by red arrows. The inset shows the coincident $\gamma$-ray spectrum gated on the summing peak located in the shaded region, where several $\gamma$-ray transitions are observed, and locations are illustrated in the corresponding level scheme \textbf{\textsl{(Right Panel)}}.
\end{figure*}

\begin{figure}[ht]
\begin{center}
\includegraphics[width=89mm]{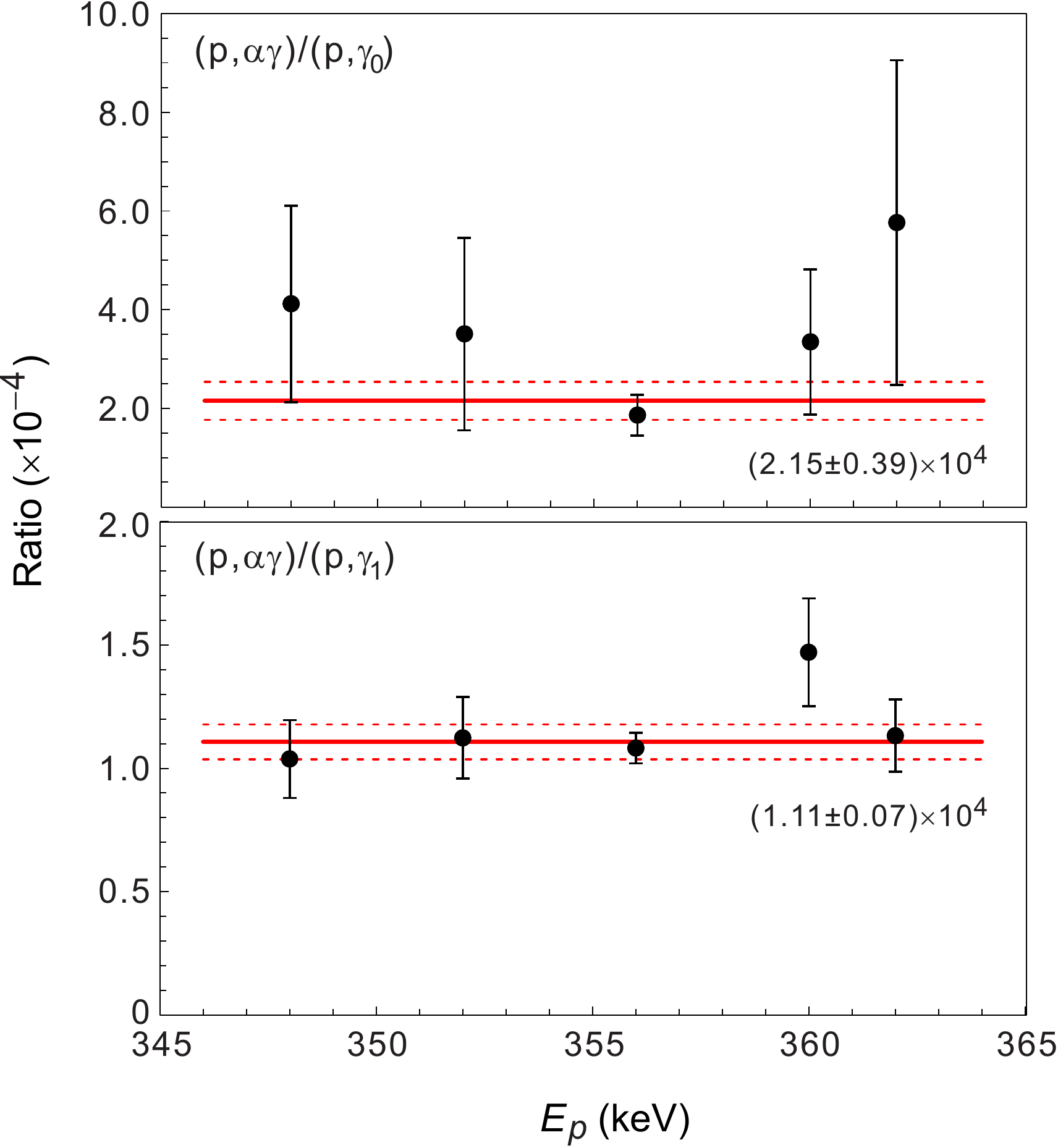}
\end{center}
\textbf{Extended Data Fig.~3: Yield ratio of ($p$, $\alpha\gamma$)/($p$, $\gamma_1$) (\textsl{Upper Panel}) and that of ($p$, $\alpha\gamma$)/($p$, $\gamma_0$) (\textsl{Lower Panel}) over the 323-keV resonance (statistical error only).} The weighted average ratios and the associated uncertainties are plotted as \textsl{solid} and \textsl{dashed lines}, respectively.
\end{figure}

\begin{figure}[ht]
\begin{center}
\includegraphics[width=183mm]{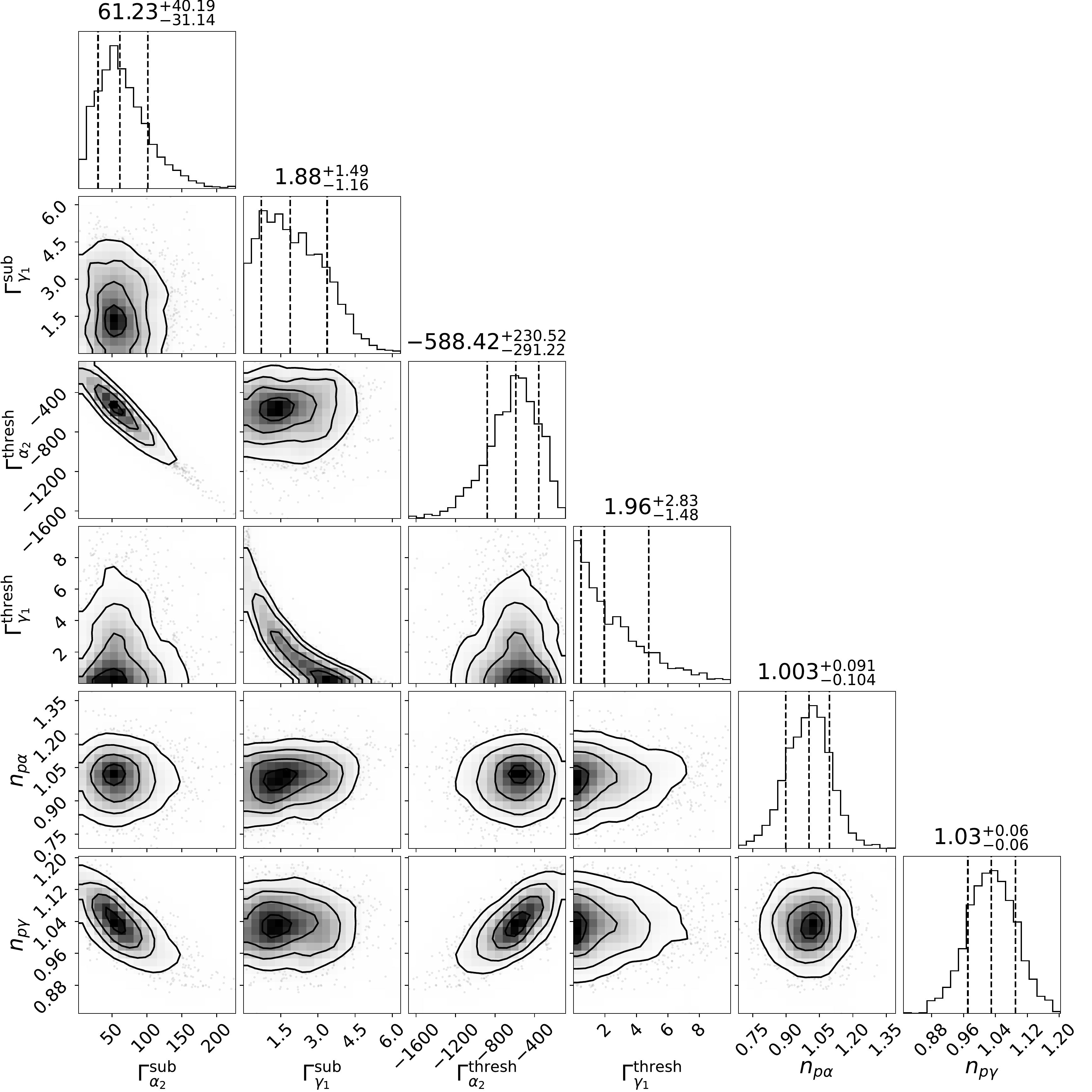}
\end{center}
\textbf{Extended Data Fig.~4. Corner plot of the covariance matrix for an MCMC analysis of the $R$-matrix fit.} The vertical dashed lines indicate 16\%, 50\%, and 84\% quantiles. Here `sub' refers to the subthreshold state at $E_x$~=~12.396~MeV, `thresh' the near threshold state at $E_x$~=~12.855~MeV, and `$n_{p\alpha}$' and `$n_{p\gamma}$' are the normalization factors for the $(p,\gamma)$ and $(p,\alpha)$ data sets, respectively. Uniform priors were taken for all parameters of the analysis. \label{fig:mcmc}
\end{figure}

\begin{table*}[h]
\footnotesize
\textbf{Extended Data Table~1. Selected astrophysical $S$ factors for $^{19}$F($p$,\,$\gamma$)$^{20}$Ne derived in this work.} The total uncertainties are listed in the parentheses, and the statistical uncertainties are listed in the last column. Conservatively, we estimate an overall systematical uncertainty of 12\%.\\ \\
%\begin{ruledtabular}
\begin{tabular}{cccc}
\hline \hline
$E_\mathrm{COM}$ (keV)   &   $S$-factor (MeV$\cdot$b) &  Statistical uncertainty (\%)\\
\hline
186.4 &	0.0140(0.0040) & 26.1   \\
195.5 &	0.0132(0.0040) & 27.4          \\
252.7 &	0.0125(0.0018) & 8.3           \\
273.8 &	0.0129(0.0028) & 18.3           \\
283.1 &	0.0262(0.0050) & 14.9           \\
291.5 &	0.0304(0.0047) & 9.7          \\
\hline \hline
\end{tabular}
%\end{ruledtabular}
\end{table*}

\begin{table*}[htp]
%\footnotesize
\textbf{Extended Data Table~2. Thermonuclear reaction rates of $^{19}$F($p$, $\gamma$)$^{20}$Ne in units of cm$^{3}$s$^{-1}$mol$^{-1}$.}
The rates are for the bare $^{19}$F nuclei in the laboratory, i.e., no thermally excited target states are considered.\\ \\
\begin{tabular}{ccccccc}
\hline \hline
%   &  \multicolumn{3}{c}{Present rate}  & NACRE~\cite{ang99}     &  deBoer21~\cite{deB21} & Williams21~\cite{wil21}\\
   &  \multicolumn{3}{c}{Present rate}  & NACRE~(\textit{3})     &  deBoer21~(\textit{5}) & Williams21~(\textit{27})\\
\cline{2-4}
$T_{9}$    &   Median  &  Low &  High &  Median & Median & Median \\
\hline
0.01  &	1.03$\times10^{-25}$ &	2.91$\times10^{-27}$ &	2.05$\times10^{-25}$ &	4.25$\times10^{-28}$ &	1.10$\times10^{-28}$ & 8.17$\times10^{-28}$ \\
0.015 &	1.30$\times10^{-21}$ &	8.78$\times10^{-23}$ &	2.52$\times10^{-21}$ &	1.37$\times10^{-23}$ &	3.50$\times10^{-24}$ & 2.61$\times10^{-23}$ \\
0.02  &	5.16$\times10^{-19}$ &	5.80$\times10^{-20}$ &	9.73$\times10^{-19}$ &	9.45$\times10^{-21}$ &	2.38$\times10^{-21}$ & 1.79$\times10^{-20}$ \\
0.03  &	9.45$\times10^{-16}$ &	1.94$\times10^{-16}$ &	1.70$\times10^{-15}$ &	3.41$\times10^{-17}$ &	8.44$\times10^{-18}$ & 6.45$\times10^{-17}$ \\
0.04  &	1.08$\times10^{-13}$ &	3.15$\times10^{-14}$ &	1.84$\times10^{-13}$ &	5.95$\times10^{-15}$ &	1.44$\times10^{-15}$ & 1.11$\times10^{-14}$ \\
0.05  &	3.13$\times10^{-12}$ &	1.16$\times10^{-12}$ &	5.09$\times10^{-12}$ &	2.34$\times10^{-13}$ &	5.57$\times10^{-14}$ & 4.38$\times10^{-13}$ \\
0.06  &	4.07$\times10^{-11}$ &	1.79$\times10^{-11}$ &	6.33$\times10^{-11}$ &	3.84$\times10^{-12}$ &	8.98$\times10^{-13}$ & 7.19$\times10^{-12}$ \\
0.07  &	3.15$\times10^{-10}$ &	1.59$\times10^{-10}$ &	4.70$\times10^{-10}$ &	3.60$\times10^{-11}$ &	8.25$\times10^{-12}$ & 6.65$\times10^{-11}$ \\
0.08  &	1.70$\times10^{-09}$ &	9.57$\times10^{-10}$ &	2.45$\times10^{-09}$ &	2.28$\times10^{-10}$ &	5.15$\times10^{-11}$ & 4.21$\times10^{-10}$ \\
0.09  &	7.15$\times10^{-09}$ &	4.39$\times10^{-09}$ &	9.91$\times10^{-09}$ &	1.10$\times10^{-09}$ &	2.46$\times10^{-10}$ & 2.02$\times10^{-09}$ \\
0.1	 &	2.53$\times10^{-08}$ &	1.69$\times10^{-08}$ &	3.39$\times10^{-08}$ &	4.31$\times10^{-09}$ &	9.93$\times10^{-10}$ & 7.76$\times10^{-09}$ \\
0.15 &	5.68$\times10^{-06}$ &	4.86$\times10^{-06}$ &	6.51$\times10^{-06}$ &	9.55$\times10^{-07}$ &	3.36$\times10^{-07}$ & 1.18$\times10^{-06}$ \\
0.2	 &	2.58$\times10^{-04}$ &	2.35$\times10^{-04}$ &	2.84$\times10^{-04}$ &	8.67$\times10^{-05}$ &	2.78$\times10^{-05}$ & 6.92$\times10^{-05}$ \\
0.3	 &	2.18$\times10^{-02}$ &	1.98$\times10^{-02}$ &	2.40$\times10^{-02}$ &	1.93$\times10^{-02}$ &	5.63$\times10^{-03}$ & 1.33$\times10^{-02}$ \\
0.4	 &	2.60$\times10^{-01}$ &	2.37$\times10^{-01}$ &	2.87$\times10^{-01}$ &	2.93$\times10^{-01}$ &	8.99$\times10^{-02}$ & 1.98$\times10^{-01}$ \\
0.5	 &	1.31$\times10^{+00}$ &	1.20$\times10^{+00}$ &	1.45$\times10^{+00}$ &	1.63$\times10^{+00}$ &	6.04$\times10^{-01}$ & 1.08$\times10^{+00}$ \\
0.6	 &	4.81$\times10^{+00}$ &	4.39$\times10^{+00}$ &	5.30$\times10^{+00}$ &	6.19$\times10^{+00}$ &	2.90$\times10^{+00}$ & 4.04$\times10^{+00}$ \\
0.7	 &	1.46$\times10^{+01}$ &	1.32$\times10^{+01}$ &	1.60$\times10^{+01}$ &	1.92$\times10^{+01}$ &	1.04$\times10^{+01}$ & 1.22$\times10^{+01}$ \\
0.8	 &	3.73$\times10^{+01}$ &	3.38$\times10^{+01}$ &	4.09$\times10^{+01}$ &	4.88$\times10^{+01}$ &	2.88$\times10^{+01}$ & 3.03$\times10^{+01}$ \\
0.9	 &	8.13$\times10^{+01}$ &	7.41$\times10^{+01}$ &	8.95$\times10^{+01}$ &	1.04$\times10^{+02}$ &	6.41$\times10^{+01}$ & 6.36$\times10^{+01}$ \\
1.0	 &	1.56$\times10^{+02}$ &	1.42$\times10^{+02}$ &	1.72$\times10^{+02}$ &	1.93$\times10^{+02}$ &	1.21$\times10^{+02}$ & 1.17$\times10^{+02}$ \\
%1.5	&	1.15$\times10^{+03}$ &	1.05$\times10^{+03}$ &	1.26$\times10^{+03}$ &	1.19$\times10^{+03}$ \\		
%2	&	3.05$\times10^{+03}$ &	2.77$\times10^{+03}$ &	3.36$\times10^{+03}$ &	2.32$\times10^{+03}$ \\		
%3	&	7.23$\times10^{+03}$ &	6.56$\times10^{+03}$ &	7.94$\times10^{+03}$ &	4.94$\times10^{+03}$ \\		
%4	&	1.00$\times10^{+04}$ &	9.12$\times10^{+03}$ &	1.10$\times10^{+04}$ &	7.75$\times10^{+03}$ \\		
%5	&	1.14$\times10^{+04}$ &	1.03$\times10^{+04}$ &	1.25$\times10^{+04}$ &	1.06$\times10^{+04}$ \\		
%6	&	1.18$\times10^{+04}$ &	1.07$\times10^{+04}$ &	1.30$\times10^{+04}$ &	1.36$\times10^{+04}$ \\		
%7	&	1.17$\times10^{+04}$ &	1.06$\times10^{+04}$ &	1.30$\times10^{+04}$ &	1.66$\times10^{+04}$ \\		
%8	&	1.13$\times10^{+04}$ &	1.03$\times10^{+04}$ &	1.25$\times10^{+04}$ &	1.97$\times10^{+04}$ \\		
%9	&	1.08$\times10^{+04}$ &	9.80$\times10^{+03}$ &	1.20$\times10^{+04}$ &	2.30$\times10^{+04}$ \\		
%10	&	1.02$\times10^{+04}$ &	9.42$\times10^{+03}$ &	1.16$\times10^{+04}$ &	2.64$\times10^{+04}$ \\	
\hline \hline
\end{tabular}
%\footnotesize
\end{table*}

\begin{table}[ht]
\textbf{Extended Data Table~3. Calcium yields (logarithm base 10 values) for fixed trajectory with \emph{constant} $\rho=39.8\,\mathrm{g}\,\mathrm{cm}^{-1}$, $T=1.19\times10^8\mathrm{K}$ and primordial initial composition~\cite{c19}.}  The trajectories were run until the hydrogen mass fraction dropped below 0.01.  Other non-degenerate binary reactions are taken from \textsc{Reaclib} v2.2.
%For the $3\alpha$ rate we use the CF88 rate~\cite{cf88}.
The reference rate for just using the original \textsc{Reaclib} rates for all binary reactions gives a mass fraction of $\log(^{40}\mathrm{Ca})=-12.33$.\\ \\
\begin{tabular}{cllrrrlrrr}
\hline \hline
\multicolumn{2}{c}{$^{19}$F($p$,\,$\gamma$) rate} & &
\multicolumn{3}{c}{NACRE $^{19}$F($p$,\,$\alpha$) rate} & &
\multicolumn{3}{c}{deBoer $^{19}$F($p$,\,$\alpha$) rate} \\
\cline{1-2}
\cline{4-6}
\cline{8-10}
&&& Low &  Mean & High && Low & Mean & High \\
\hline \hline
\multirow{3}{3.5em}{JUNA} & Low && -11.59 & -11.72 & -11.83 && -11.70 & -11.68 & -11.90 \\
& Mean && -11.47 & -11.60 & -11.70 && -11.57 & -11.55 & -11.77 \\
& High && -11.37 & -11.50 & -11.60 && -11.48 & -11.46 & -11.67 \\
\noalign{\medskip}
\multirow{3}{3.5em}{NACRE} & Low && -12.51 & -12.64 & -12.75 && -12.62 & -12.60 & -12.82 \\
& Mean && -12.21 & -12.35 & -12.45 && -12.32 & -12.30 & -12.52 \\
& High && -12.04 & -12.17 & -12.27 && -12.15 & -12.13 & -12.34 \\
\noalign{\medskip}
\multirow{3}{3.5em}{deBoer} & Low && -13.25 & -13.38 & -13.48 && -13.36 & -13.34 & -13.55 \\
& Mean && -12.81 & -12.94 & -13.05 && -12.92 & -12.90 & -13.12 \\
& High && -12.05 & -12.18 & -12.28 && -12.16 & -12.13 & -12.35 \\
\noalign{\medskip}
\multirow{3}{3.5em}{Williams} & Low && -12.07 & -12.21 & -12.31 && -12.18 & -12.16 & -12.38 \\
& Mean && -11.99 & -12.13 & -12.23 && -12.10 & -12.08 & -12.30 \\
& High && -11.91 & -12.04 & -12.15 && -12.02 & -12.00 & -12.22 \\
\hline \hline
\end{tabular}
\end{table}

\begin{table}[ht]
\textbf{Extended Data Table~4. Similar to Extended Data Table~3 but uses the actual central temperature trajectory of a 40~$\mathrm{M}_\odot$ Population III star model~\cite{hw10} and using a mixing model to emulate convection.}  The trajectory was run until a core hydrogen mass fraction of $0.01$.  The mixing model assumes that the trajectory represents a fraction of $\taumix=0.0595$ of the total reservoir such that burning reduced the H mass fraction in the trajectory to $0.01$, consistent with the stellar model.  The reference rate for just using the original \textsc{Reaclib} rates for all binary reactions is $\log(^{40}\mathrm{Ca})=-12.02$.\\ \\
\begin{tabular}{cllrrrlrrr}
\hline \hline
\multicolumn{2}{c}{$^{19}$F($p$,\,$\gamma$) rate} & &
\multicolumn{3}{c}{NACRE $^{19}$F($p$,\,$\alpha$) rate} & &
\multicolumn{3}{c}{deBoer $^{19}$F($p$,\,$\alpha$) rate} \\
\cline{1-2}
\cline{4-6}
\cline{8-10}
&&& Low &  Mean & High && Low & Mean & High \\
\hline \hline
\multirow{3}{3.5em}{JUNA} & Low && -11.29 & -11.44 & -11.54 && -11.36 & -11.41 & -11.66 \\
& Mean && -11.14 & -11.29 & -11.39 && -11.22 & -11.26 & -11.51 \\
& High && -11.03 & -11.18 & -11.29 && -11.11 & -11.15 & -11.41 \\
\noalign{\medskip}
\multirow{3}{3.5em}{NACRE} & Low &&  -12.19 & -12.33 & -12.44 && -12.26 & -12.31 & -12.56\\
& Mean && -11.89 & -12.04 & -12.14 && -11.97 & -12.01 & -12.26 \\
& High && -11.72 & -11.86 & -11.97 && -11.79 & -11.83 & -12.09 \\
\noalign{\medskip}
\multirow{3}{3.5em}{deBoer} & Low && -13.13 & -13.27 & -13.38 && -13.20 & -13.24 & -13.49 \\
& Mean && -12.52 & -12.66 & -12.77 && -12.59 & -12.63 & -12.89 \\
& High && -11.72 & -11.86 & -11.97 && -11.79 & -11.83 & -12.09 \\
\noalign{\medskip}
\multirow{3}{3.5em}{Williams} & Low && -11.74 & -11.88 & -11.99 && -11.81 & -11.85 & -12.11 \\
& Mean && -11.66 & -11.80 & -11.91 && -11.73 & -11.77 & -12.03 \\
& High && -11.57 & -11.71 & -11.82 && -11.64 & -11.69 & -11.94 \\
\hline \hline
\end{tabular}
\end{table}

\begin{table}[ht]
\textbf{Extended Data Table~5. Calcium yields (logarithm base 10 values) for full stellar models.}  The first data column gives the average abundance over the entire star at the terminal-age main-sequence (TAMS), defined by a core hydrogen mass fraction of $0.01$, consistent with what we use elsewhere.  The last three columns list the average calcium mass fraction at the pre-supernova stage: in the hydrogen envelope (hydrogen mass fraction $\ge0.01$), in the helium shell (helium mass fraction $\ge0.01$ and hydrogen mass fraction $<0.01$), and in the combination of both (helium mass fraction $\ge0.01$), respectively. \\ \\
%The last row gives the historic reference using all rates from~\cite{hw10} whereas the other rows used \textsc{Reaclib} 2.2 as base reaction rate set except $^{19}$F(p,$\gamma$) and $^{19}$F(p,$\alpha$) as listed.
\begin{tabular}{llllllrlrrr}
\hline \hline
\multicolumn{2}{c}{$^{19}$F($p$,\,$\gamma$) rate} & &
\multicolumn{2}{c}{$^{19}$F($p$,\,$\alpha$) rate} & &
\multicolumn{1}{c}{TAMS} & &
\multicolumn{3}{c}{Pre-supernova} \\
\cline{1-2}
\cline{4-5}
\cline{7-7}
\cline{9-11}
&&&&&& \multicolumn{1}{c}{Star} && H envel. & He shell & H+He envel. \\
\hline \hline
\multirow{3}{3.5em}{JUNA} & Low && \multirow{3}{3.5em}{NACRE} & High && -10.81 && -11.18 & -10.74 & -11.13 \\
& Mean &&& Mean && -11.10 && -11.47 & -11.05 &-11.41 \\
& High &&& Low  && -11.41 && -11.77 & -11.39 &-11.73 \\
\noalign{\medskip}
\multirow{3}{3.5em}{NACRE} & Low && \multirow{3}{3.5em}{NACRE} & High && -11.56 && -11.93 & -11.51 & -11.88 \\
& Mean &&& Mean && -11.90 && -12.27 & -11.86 & -12.21 \\
& High &&& Low  && -12.31 && -12.68 & -12.25 &-12.63 \\
\noalign{\medskip}
\multirow{3}{3.5em}{deBoer} & Low && \multirow{3}{3.5em}{NACRE} & High && -11.56 && -11.92 & -11.49 & -11.88 \\
& Mean &&& Mean && -12.51 && -12.88 & -12.46 & -12.85 \\
& High &&& Low  && -13.62 && -13.98 & -13.57 & -13.94 \\
\noalign{\medskip}
\multirow{3}{3.5em}{Williams} & Low && \multirow{3}{3.5em}{NACRE} & High && -11.40 && -11.77 & -11.39 & -11.73 \\
& Mean &&& Mean && -11.64 && -12.01 & -11.62 & -11.96 \\
& High &&& Low  && -11.83 && -12.20 & -11.83 & -12.16 \\
\noalign{\smallskip}
%\hline
%\noalign{\smallskip}
%\multicolumn{5}{c}{--- HW10 ---} && -9.88 && -10.22 & -9.87 &-10.17\\
\hline \hline
\end{tabular}
\end{table}

\end{document}